\documentclass[10pt,conference]{IEEEtran}
\IEEEoverridecommandlockouts
\usepackage{cite}
\usepackage{amsmath,amssymb,amsfonts}
\usepackage{algorithmic}
\usepackage{graphicx}
\usepackage{textcomp}
\usepackage{xcolor}
\usepackage{algorithm}
\usepackage{algorithmic}
\usepackage{setspace}
\usepackage[T1]{fontenc}
\def\BibTeX{{\rm B\kern-.05em{\sc i\kern-.025em b}\kern-.08em
    T\kern-.1667em\lower.7ex\hbox{E}\kern-.125emX}}
\usepackage{enumerate}
\usepackage{hyperref}
\hypersetup{hidelinks}

\begin{document}


\title{Flexible Bidding in Service-Oriented Combinatorial Spectrum Forward Auctions}



\author{Xiang Shao$^{\ast}$, Wei Wang$^\dag$, and Guan Gui$^{\ddag}$ \\~\\	
	$^{\ast}$College of Electronic and Information Engineering, NUAA, Nanjing, China\\
	$^{\dag}$School of Information and Communication Engineering, Xi'an Jiaotong University, Xi'an, China\\
	$^\ddag$College of Telecommunications and Information Engineering, NJUPT, Nanjing, China\\
	Email: XiangShao1226@nuaa.edu.cn, w25wang@xjtu.edu.cn, guiguan@njupt.edu.cn}



\maketitle

\begin{abstract}

Traditional combinatorial spectrum auctions mainly rely on fixed bidding and matching processes, which limit participants' ability to adapt their strategies and often result in suboptimal social welfare in dynamic spectrum sharing environments. To address these limitations, we propose a novel approximately truthful combinatorial forward auction scheme with a flexible bidding mechanism aimed at enhancing resource efficiency and maximizing social welfare. In the proposed scheme, each buyer submits a combinatorial bid consisting of the base spectrum demand and adjustable demand ranges, enabling the auctioneer to dynamically optimize spectrum allocation in response to market conditions. To standardize the valuation across heterogeneous frequency bands, we introduce a Spectrum Equivalent Mapping (SEM) coefficient. A greedy matching algorithm is employed to determine winning bids by sorting buyers based on their equivalent unit bid prices and allocating resources within supply constraints. Simulation results demonstrate that the proposed flexible bidding mechanism significantly outperforms existing benchmark methods, achieving notably higher social welfare in dynamic spectrum sharing scenarios.

\end{abstract}


\section{Introduction}

Future sixth-generation (6G) wireless networks are envisioned to deliver unprecedented user experiences across a wide range of application scenarios, driving the integration of a broad spectrum that includes sub-6 GHz, millimeter wave (mmWave), and terahertz (THz) frequency bands \cite{Wang2023}.
While high-frequency bands have garnered increasing research attention, efficient utilization of mid-band spectrum remains crucial due to its optimal balance between coverage and capacity \cite{You2021}.
To enhance overall spectrum efficiency, spectrum sharing strategies have become indispensable, particularly those involving mobile network operators (MNOs) leasing underutilized spectrum to vertical industries or secondary operators lacking dedicated licensed bands \cite{Shao2023}.

Given the diversity of service requirements in 6G, such as enhanced Mobile Broadband (eMBB), which demands high bandwidth, and Ultra-Reliable Low-Latency Communications (uRLLC), which prioritizes reliability and minimal latency over spectrum volume, a one-size-fits-all approach to spectrum allocation is no longer sufficient.
Consequently, operators must aggregate heterogeneous frequency bands rather than relying on a single band. Combinatorial auctions provide a promising solution by allowing buyers to bid on indivisible bundles of multiple frequency bands, tailored to specific service needs \cite{ref22}. \looseness=-1

The heterogeneous valuations of different frequency bands may incentivize buyers to submit dishonest bids on individual items within a bundle to maximize private utility, thereby undermining the incentive for truthful bidding \cite{ref22}. To maintain truthfulness in combinatorial spectrum auctions, most existing approaches adopt a bundle-level pricing scheme rather than itemized pricing, which prevents bid manipulation. For example, the approach in \cite{CAICC} computed only an aggregate price for each bidder's package rather than separately pricing the constituent wireless resource blocks and processing units. Similarly, \cite{LiChangle2014} introduced a two-dimensional combinatorial auction for time-frequency spectrum resources, maintaining truthfulness by pricing the entire time-frequency bundle as a single entity rather than its individual components.

However, in scenarios involving heterogeneous frequency bands, such rigid package-level mechanisms may lead to inefficient allocations and reduced social welfare. This is primarily due to conflicts among overlapping bidding packages, especially when the auction mechanism lacks sufficient flexibility.
For instance, \cite{Zhaohaitao2025} employed an \textit{all-or-nothing} allocation approach where wireless and computing resources in each package are either fully allocated or entirely rejected, significantly reducing resource utilization efficiency. To enhance auction flexibility, \cite{Dong2021} introduced the THIMBLE mechanism, allowing users to submit multiple virtual group bids that capture diverse channel requirements. Nevertheless, the THIMBLE mechanism only considers homogeneous frequency bands, limiting its applicability to heterogeneous spectrum combinations.
Furthermore, while increased flexibility enables buyers to adapt their bids better to dynamic market conditions, it may undermines truthfulness due to increased manipulation opportunities. 
These challenges highlight the need for a spectrum combinatorial auction framework that effectively balances allocation flexibility with incentives for truthful bidding.

In this paper, we propose an approximately truthful flexible combinatorial forward auction for dynamic spectrum sharing involving multiple buyers and a single seller. Each buyer submits a combinatorial bid comprising its demand for various frequency bands along with adjustable ranges, bundled with a bidding price.
The seller, in turn, declares its supply of multiple frequency bands and a reserved unit price.
To enable flexible and fair spectrum allocation across heterogeneous frequency bands, we introduce a Spectrum Equivalence Mapping (SEM) coefficient that standardizes different frequency bands into a unified equivalent spectrum space.
The main contributions of this paper are summarized as follows:
\begin{itemize}

	\item To overcome the rigidity of fixed bidding while preserving truthful behavior, we propose an approximately truthful combinatorial forward auction that allows buyers to specify both base demands and adjustable ranges. The introduction of the SEM coefficient enables flexible substitution among different frequency bands within these ranges.

	\item We design an efficient Greedy Matching-based Winner Determination (GMWD) mechanism that sorts buyers by their equivalent unit bid prices in descending order and allocates spectrum resources until supply constraints are met. We theoretically prove that the proposed mechanism satisfies key economic properties, including approximate truthfulness, individual rationality, and budget balance.

	\item Extensive simulation results demonstrate that proposed auction mechanism achieves superior social welfare compared to two benchmark auction schemes, validating the effectiveness of flexible bidding. We also analyze the impact of adjustment range parameters under different system configurations, demonstrating the robustness and adaptability of our approach.
	\end{itemize}

\section{System Model}

We consider a secondary spectrum auction market in a defined geographical region, where a primary operator (PO) leases available spectrum resources to multiple secondary operators (SOs) who require spectrum to serve their subscribers \cite{Scenario}. The auctioned spectrum resources are uniformly partitioned across both spatial and temporal dimensions\footnote{The auctioned spectrum units cover identical time periods but vary in frequency, with durations significantly shorter than those in primary spectrum markets}\cite{ShaoXiangIoTJ}. Each bidder combines frequency bands to meet specific service requirements. In this framework, the PO serves as the seller denoted as $SE$, $M$ SOs participate as buyers denoted as $\{BU_1, BU_2, \ldots, BU_M\}$, and an authoritative auctioneer facilitates the process. The auction involves $K$ distinct types of frequency bands.

The SEM coefficient is defined as
\begin{equation}
	\boldsymbol{\rho}=\{\rho^1, \rho^2, \ldots, \rho^K\},
\end{equation}
where $\rho^k$ represents the conversion ratio between frequency band $k$ and the equivalent public band space. This coefficient establishes a unified valuation metric across different frequency bands in the auction.

The bid from buyer $BU_m$ is expressed as
\begin{equation}
	\mathbf{D_m}=\left( \textless D_m^1, \Delta D_m^1\textgreater,\ldots, \textless D_m^K, \Delta D_m^K\textgreater, b_m\right),
	\label{bidding}
\end{equation}
where $D_m^k$ is the base bid denoting the requested quantity of frequency band $k$, $\Delta D_m^k$ represents the adjustable bandwidth that can be substituted with other bands, and $b_m$  indicates the bid price for package $\mathbf{D_m}$.

The seller's spectrum offering is characterized by
\begin{equation}
	\mathbf{L}=\left(L^1, \ldots, L^K, r\right),
	\label{asking}
\end{equation}
where $L^k$ is the available quantity of frequency band $k$ and $r$ denotes the reserved unit price.

We assume that the total spectrum demand significantly exceeds the available supply in the considered scenario \cite{Mochaourab}. The all-or-nothing allocation scheme requires that buyers either receive their complete requested package or nothing at all. The auction is designed to maximize social welfare, defined as the difference between the winning buyers' total bid prices and the reserve price of allocated spectrum, formulated as
\begin{subequations}
	\begin{align}
		&\max_{ \{x_m\},\{y_m^k\} } \sum_{m=1}^M x_m b_m -  \sum_{m=1}^M x_m \left( \sum_{k=1}^K D^k_m \rho^k r\right) \label{0} \\
		&\textbf{s.t.}\quad \sum_{m=1}^M x_m \left(D_m^k-y_m^k \right) \leq L^k, \forall k, \label{c1} \\
		&\quad\quad \;\; \sum_{k=1}^K\sum_{m=1}^M  x_m \rho^k D_m^k \leq \sum_{k=1}^K\rho^k L^k, \label{c3} \\
		& \quad\quad \;\; 0\leq y_m^k\leq \Delta D_m^k, \forall m, \forall k, \label{c2} \\
		&\quad \quad \;\;  x_m \in \{0,1\}, \forall m. \label{c4}
			\end{align}
\end{subequations}
This is a mixed-integer programming problem. The binary variable $\{x_m\}$ indicates winning packages, while the continuous variable $\{y_m^k\}$ represents spectrum adjustment ranges in the allocation outcome. The constraints are defined as follows: (i) Constraint \eqref{c1} ensures adjusted frequency bands comply with specified band quantities; 
(ii) Constraint \eqref{c3} guarantees that equivalent public spectrum does not exceed the available supply;
(iii) Constraint \eqref{c2} enforces that adjustments remain within buyers' permitted ranges.
To solve this problem, we employ the SEM coefficient to sort buyers by their equivalent unit bidding prices in descending order. A greedy matching-based winner determination (GMWD) mechanism then selects winning bids and optimizes spectrum allocation.

\section{Proposed Greedy Matching-based Winner Determination Mechanism and Payment}
\subsection{GMWD Mechanism}

We propose a GMWD mechanism that prioritizes buyers based on their equivalent unit bidding prices in descending order. The mechanism sequentially processes buyers according to this ranking to determine winners and optimize spectrum allocation.

The equivalent unit bidding price in the public spectrum for buyer $BU_m$ is calculated as
\begin{equation}
	b_m^{\text{eq}}=\frac{b_m}{\sum_{k=1}^K D^k_m \rho^k }.
	\label{equivalent}
\end{equation}
We then resort buyers by their equivalent unit prices in descending order
\begin{equation}
	S=\{b_{1^*}^{\text{eq}}, b_{2^*}^{\text{eq}}, \ldots, b_{M^*}^{\text{eq}} \}.
\end{equation}
The selection process proceeds from the highest-ranked buyer in $S$ until the available spectrum resources are exhausted. When frequency band constraints are violated, the SEM coefficient facilitates substitution of spectrum within buyers' adjustment ranges with alternative frequency bands. To ensure the maximal social welfare, we set $y_m^k=\Delta D_m^k, \forall m, \forall k$ before the mechanism starts. The GMWD mechanism is detailed in Algorithm \ref{GMWD}.

\begin{algorithm}[htb] 
	\caption{Greedy Matching Winner Determination (GMWD) Mechanism} 
	\label{GMWD} 
	\begin{algorithmic}[1]
		\renewcommand{\algorithmicrequire}{\textbf{Input:}}
		\REQUIRE
		Buyers sort $S$, Spectrum information $\mathbf{L}$, Buyers bids $\mathbf{D_m}$, SEM coefficient $\boldsymbol{\rho}$;\\
		\renewcommand{\algorithmicensure}{\textbf{Output:}}
		\ENSURE
		$\{x_m\}$;
		\STATE Find buyers in $S$ whose $b_{m^*}^{\text{eq}} \geq r$ as $\{b_{1^*}^{\text{eq}}, \ldots, b_{M'}^{\text{eq}}\}$;
		\STATE Set the equivalent remaining spectrum as\\ $ES=\sum_{k=1}^K L^k \rho^k$;
		\STATE Set the actual remaining spectrum as\\ $AS=\{ L^1, \ldots, L^K \}$;
		\FOR{buyer $m^*=1$ to $M'$}
		\STATE Renew the actual remaining spectrum \\
		$AS \gets AS-\{D^1_{m^*}-\Delta D^1_{m^*}, \ldots, D^K_{m^*}- \Delta D^K_{m^*}\}$;
		\STATE Renew the equivalent remaining spectrum\\
		 $ES \gets ES- \sum_{k=1}^K D^k_{m^*} \rho^k $;
		 \IF {any element in $AS$ is less than $0$ or $ES< 0$}
		 \STATE{\textbf{Break} \textbf{for}};
		 \ELSE
		 \STATE $x_{m^*}=1$;		 
		 \ENDIF
		\ENDFOR
		\RETURN $\{x_m\}$.
	\end{algorithmic}
\end{algorithm}

\subsection{Payment Determination}

Let $N$ represents the number of winning buyers, with winning buyers denoted as $\{BU_{1^*}, BU_{2^*}, \ldots, BU_{N}\}$. The payment structure for winning buyers follows two distinct cases:
\begin{enumerate}[1)]
	\item If $b_{N}^{\text{eq}}\geq r$ and $b_{N+1}^{\text{eq}}\geq r$ in $S$, the payment of winning buyer $BU_{m^*}$ is $b_{N+1}^{\text{eq}}\left( \sum_{k=1}^K D_{m^*}^k \rho^k\right)$;
	\item If $b_{N}^{\text{eq}}\geq r$ and $b_{N+1}^{\text{eq}} \leq r$ in $S$, the payment of winning buyer $BU_{m^*}$ is $r\left( \sum_{k=1}^K D_{m^*}^k \rho^k\right)$.
\end{enumerate}
The seller receives payment based on the total equivalent spectrum sold at the reserved unit price, which denoted as $r \big(\sum_{m^*=1^*}^{N}\sum_{k=1}^K D^k_{m^*} \rho^k \big)$.

\section{Economic Properties Analyses}

 This section analyzes the economic properties of the proposed mechanism, including approximately truthfulness, individual rationality, and budget balance. As our combinatorial auction follows a forward auction model where only buyers submit bids, the mechanism must guarantee approximate truthfulness for buyers while maintaining individual rationality and budget balance for all participants.

\textbf{ Proposition 1:} GMWD mechanism is approximately truthful (in the sense of hard to manipulate, HTM) for all buyers.

\textbf{Proof:} According to \cite{Wang2014}, an approximately truthful auction can be either \textit{truthful in expectation} or \textit{hard to manipulate} (HTM). In this paper, we focus on the latter.

\textbf{Definition 1 (Hard to Manipulate, HTM):} A mechanism is HTM if, for any buyer $BU_m$, determining whether there exists a misreport $\mathbf{\tilde{D}_m}$ that increases her utility by at least any $\varepsilon>0$ is NP-hard.

Thus, to prove that GMWD is approximately truthful, we only need to establish that it is HTM, leading to the \textbf{following proposition 2}.

\textbf{Proposition 2:} GMWD mechanism is HTM.

\textbf{Proof:} The proof is conducted in two steps:

\textbf{Step 1 (Structural Properties):} From Sanghvi \& Parkes \cite{Sanghvi2004},
a forward combinatorial auction mechanism that meets the following two structural conditions is known to be HTM:
\begin{itemize}
	\item \textbf{Greedy Optimality (G-OPT):}  The mechanism never leaves requested spectrum idle and never allocates beyond any buyer's request.
	\item \textbf{Strong Consumer Sovereignty (SCS):} No single buyer's bid can exceed the final realized social welfare.
\end{itemize}
We first verify these \textbf{two conditions} for GMWD:

\textbf{Condition 1 (GMWD satisfies G-OPT):} GMWD scans buyers in descending order of their valuations $b_m^{\text{eq}}$. A buyer is allocated spectrum only if:
\begin{itemize}
	\item Every band's residual capacity $AS$ after deducting its non-adjustable allocation (line 5) is non-negative.
	\item The equivalent budget $ES$ remains non-negative (line 6).
\end{itemize}
At termination:
\begin{itemize}
	\item Either some band is fully allocated, or the equivalent budget is less than zero.
	\item Thus, no unallocated spectrum remains that can fulfill any losing buyer’s request, and no winner receives more than requested. 
	\end{itemize}

\textbf{Condition 2 (GMWD satisfies SCS):}
Let $SW$ denote the realized social welfare. Suppose that there is a buyer $BU_m$ with a bid $b_m>SW$:
\begin{itemize}
	\item If $BU_m$ loses, their higher bid would prioritize them at the top of the sorted list, contradicting the loss.
	\item If $BU_m$ wins, then their bid is already included in $SW$, contradicting $b_m>SW$.
\end{itemize}
Thus, no such buyer exists, confirming that GMWD meets SCS.

\textbf{Step 2 (Reduction to NP-hard Problem):}
To rigorously demonstrate the computational hardness of buyer manipulation, we first reduce the general Winner Determination (WD) problem to a simplified yet representative NP-hard problem.
We now show that deciding profitable manipulation is computationally intractable, even in a simplified unit-band scenario.

We consider a special case with:
\begin{itemize}
	\item A single band ($K=1$), no adjustments ($\Delta D^1_m=0$);
	\item Seller capacity $L_1$, buyers' demands $D_m^1$, and bids $b_m$.
\end{itemize}
The WD problem which selects a subset of buyers to maximize $\sum b_m$ under constraint $\sum D_m^1 \leq L^1$, is equivalent to the classical 0/1 Knapsack problem, known to be NP-complete. Hence, even this restricted WD problem is NP-hard. 

In summary, the proposed mechanism is approximately truthful. $\hfill\blacksquare$

\textbf{Proposition 3:} The GMWD mechanism satisfies individual rationality for all participants.

\textbf{Proof:} For any losing buyer, the utility is zero. For a winning buyer $BU_{m^*}$, the utility is non-negative in both cases:
\begin{equation}
		b_{m^*}-b_{N+1}^{\text{eq}}\big( \sum_{k=1}^K D_{m^*}^k \rho^k\big)=\big( b_{m^*}^{\text{eq}}-b_{N+1}^{\text{eq}}\big) \big( \sum_{k=1}^K D_{m^*}^k \rho^k\big)\geq 0, \nonumber
\end{equation}
or
\begin{equation}
		b_{m^*}-r\big( \sum_{k=1}^K D_{m^*}^k \rho^k\big)=\big(  b_{m^*}^{\text{eq}}-r \big) \big( \sum_{k=1}^K D_{m^*}^k \rho^k\big)\geq 0. \nonumber
	\end{equation}
The seller's utility is zero since the payment equals the reserve price. Thus, all participants have non-negative utilities, proving individual rationality. $\hfill\blacksquare$

\textbf{Proposition 4:} The GMWD mechanism maintains budget balance for the auctioneer.

\textbf{Proof:} The auctioneer's utility is
\begin{subequations}
		\begin{align}
	\nonumber	&\sum_{m^*=1^*}^{N} b_{m^*}-r\big(\sum_{m^*=1^*}^{N}\sum_{k=1}^K D^k_{m^*} \rho^k \big)\\ \nonumber
		=&\sum_{m^*=1^*}^{N} b_{m^*}^{\text{eq}} \big( \sum_{k=1}^K D_{m^*}^k \rho^k\big)- r\big(\sum_{m^*=1^*}^{N}\sum_{k=1}^K D^k_{m^*} \rho^k \big)\geq 0. \nonumber
		\end{align}
	\end{subequations}

The non-negativity of this expression confirms the budget balance property. $\hfill\blacksquare$

\section{Simulation and Discussion}

This section presents simulation results evaluating the performance of the proposed flexible bidding mechanism. We consider a system with $K = 5$ frequency bands. The seller's available channels per band follow a Poisson distribution \cite{TVT2022} with mean values uniformly distributed in $[50, 100]$, while buyer demands follow a Poisson distribution with means in $[8, 16]$.
All buyers share identical adjustment ranges ($\Delta$) across frequency bands.
The SEM coefficients $\boldsymbol{\rho}$ are set as $\{10,8,6,4,2\}$.
For comparison, we implement two benchmark mechanisms: 1) The TCDA mechanism \cite{TCDA} employing binary $\{0,1\}$ package allocation; 2) The THIMBLE mechanism \cite{Dong2021} utilizing virtual bids for enhanced flexibility.
Each configuration undergoes 10,000 Monte Carlo simulations.

\begin{figure}
	\centering
	\includegraphics[width=0.34\textwidth]{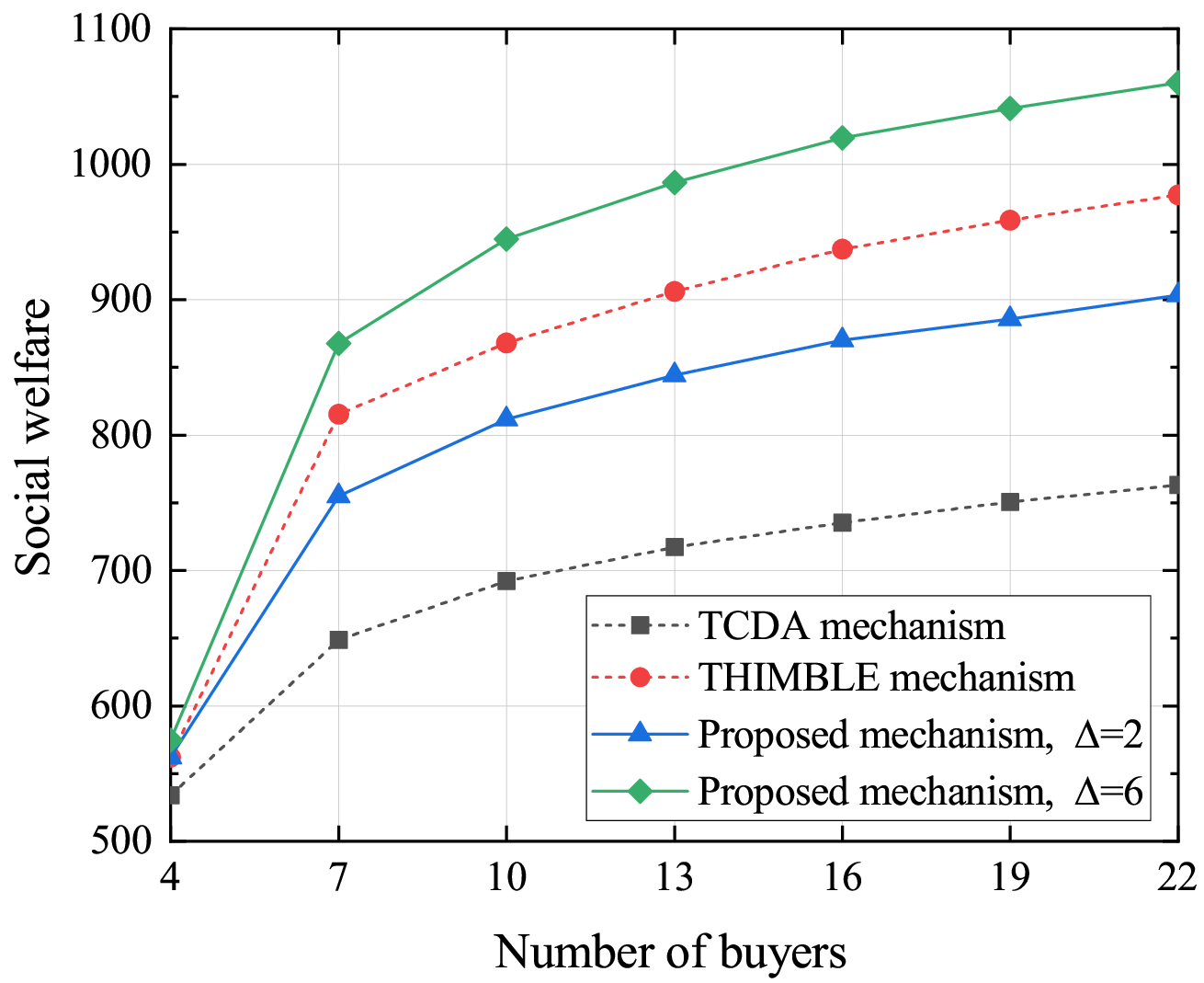}
	\caption{Social welfare comparison under different number of buyers.}
	\label{Algorithm comparison}
\end{figure}

We first examine how the number of buyers affects the social welfare with different auction mechanisms to validate whether our proposed flexible bidding mechanism consistently outperforms benchmarks as competition increases.
 As illustrated in Fig. \ref{Algorithm comparison}, the proposed mechanism with $\Delta = 6$ consistently achieves the highest social welfare, outperforms both TCDA and THIMBLE benchmarks. The configuration with $\Delta= 2$ provides lower social welfare than THIMBLE due to its relatively limited flexibility, as THIMBLE leverages additional flexibility through virtual bids. The TCDA mechanism yields the lowest performance, highlighting the inefficiency of the \textit{all-or-nothing} allocation in traditional combinatorial auctions.

\begin{figure}
	\centering
	\includegraphics[width=0.34\textwidth]{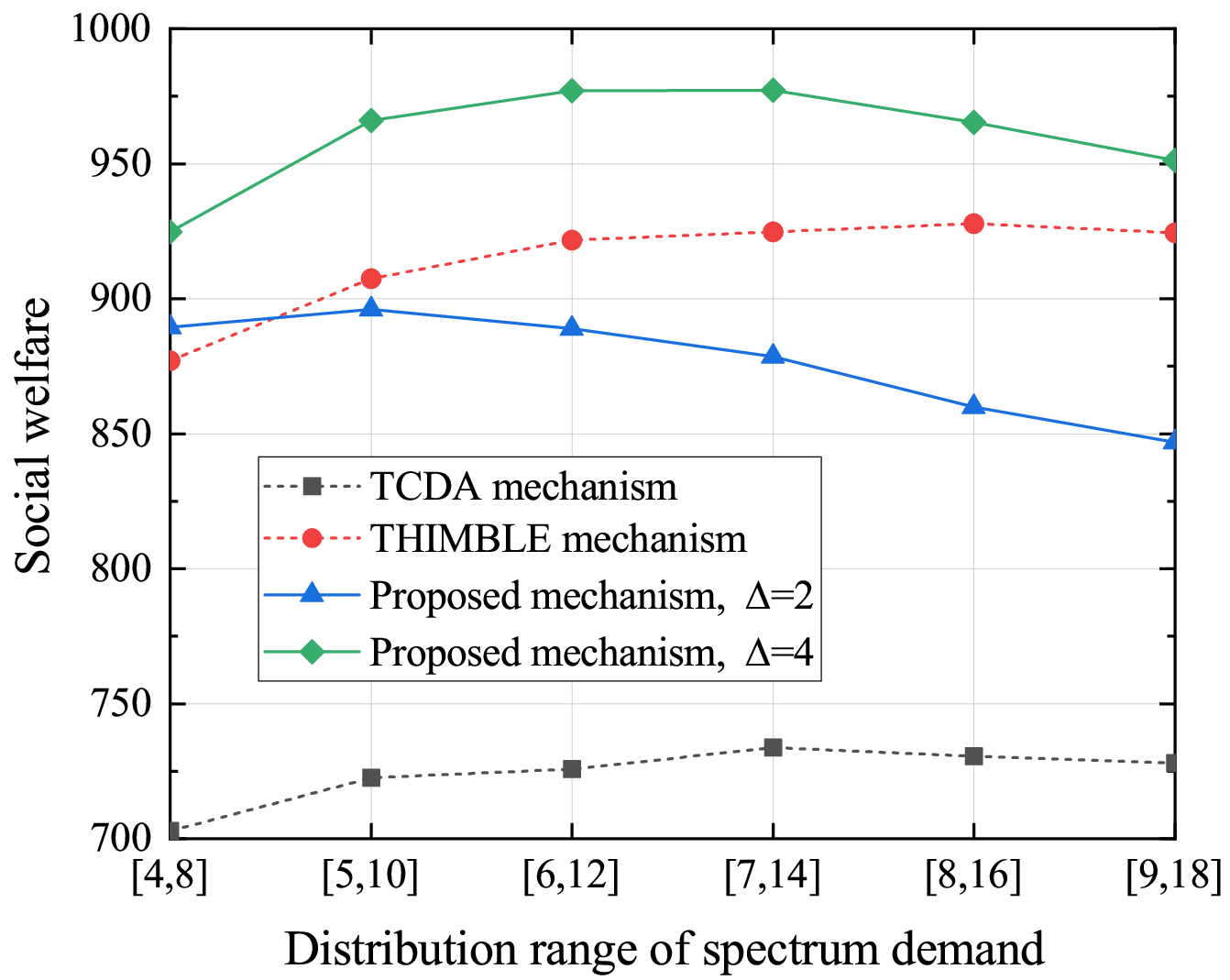}
	\caption{Social welfare comparison under different spectrum demand distribution ranges, where the number of buyers is 15.}
	\label{distribution}
\end{figure}

We next evaluate how varying buyer spectrum demand distributions impacts social welfare and test whether larger adjustment ranges enhance performance under diverse bid combinations. 
As depicted in Fig. \ref{distribution}, when the range of the buyers' spectrum demand distribution expands, the bid combinations become more diverse.
Initially, social welfare with the proposed mechanism with $\Delta=2$ and $4$ improves but subsequently declines, with the turning point for $\Delta=2$ appearing earlier compared to $\Delta=4$.
This demonstrates that a larger adjustment range enhances the proposed mechanism's capability to accommodate diverse bid combinations, maintaining high social welfare even under highly varying demand scenarios. The TCDA mechanism, without sufficient flexibility, consistently achieves the lowest social welfare.

\begin{figure}
	\centering
	\includegraphics[width=0.34\textwidth]{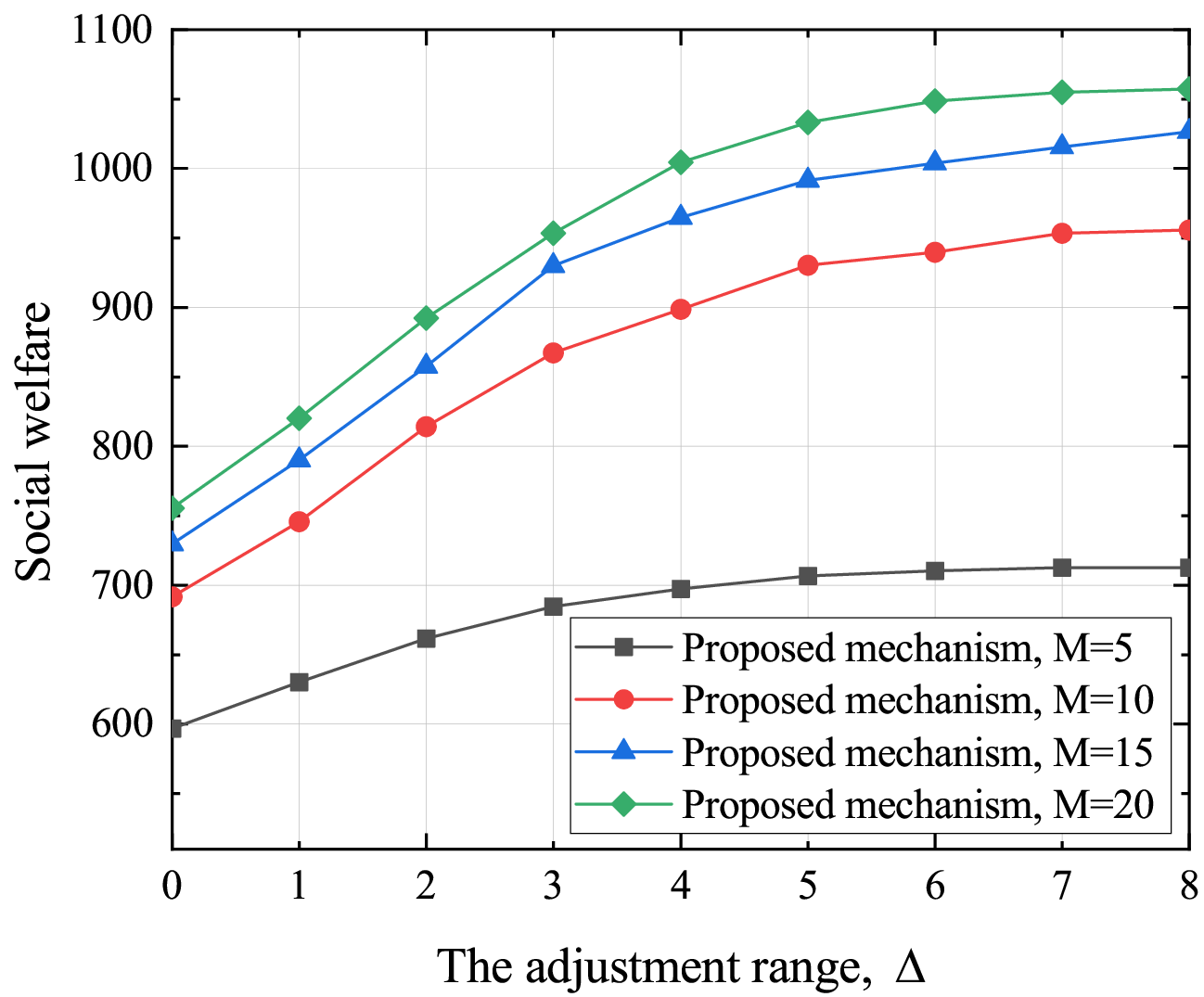}
	\caption{Social welfare comparison under different adjustment ranges, where the number of buyers is 5, 10, 15 and 20, respectively.}
	\label{varying range}
\end{figure}

To highlight the importance of flexible adjustment ranges, we examine their direct effect on social welfare across different number of buyers.
 As shown in Fig. \ref{varying range}, increasing the adjustment range notably improves social welfare with our proposed mechanism, and increasing buyers will increase competitive bidding and thus the social welfare. The clear upward trend across varying buyer counts and adjustment ranges underscores the importance of flexible bidding in maximizing spectrum utilization efficiency.

\begin{figure}
	\centering
	\includegraphics[width=0.34\textwidth]{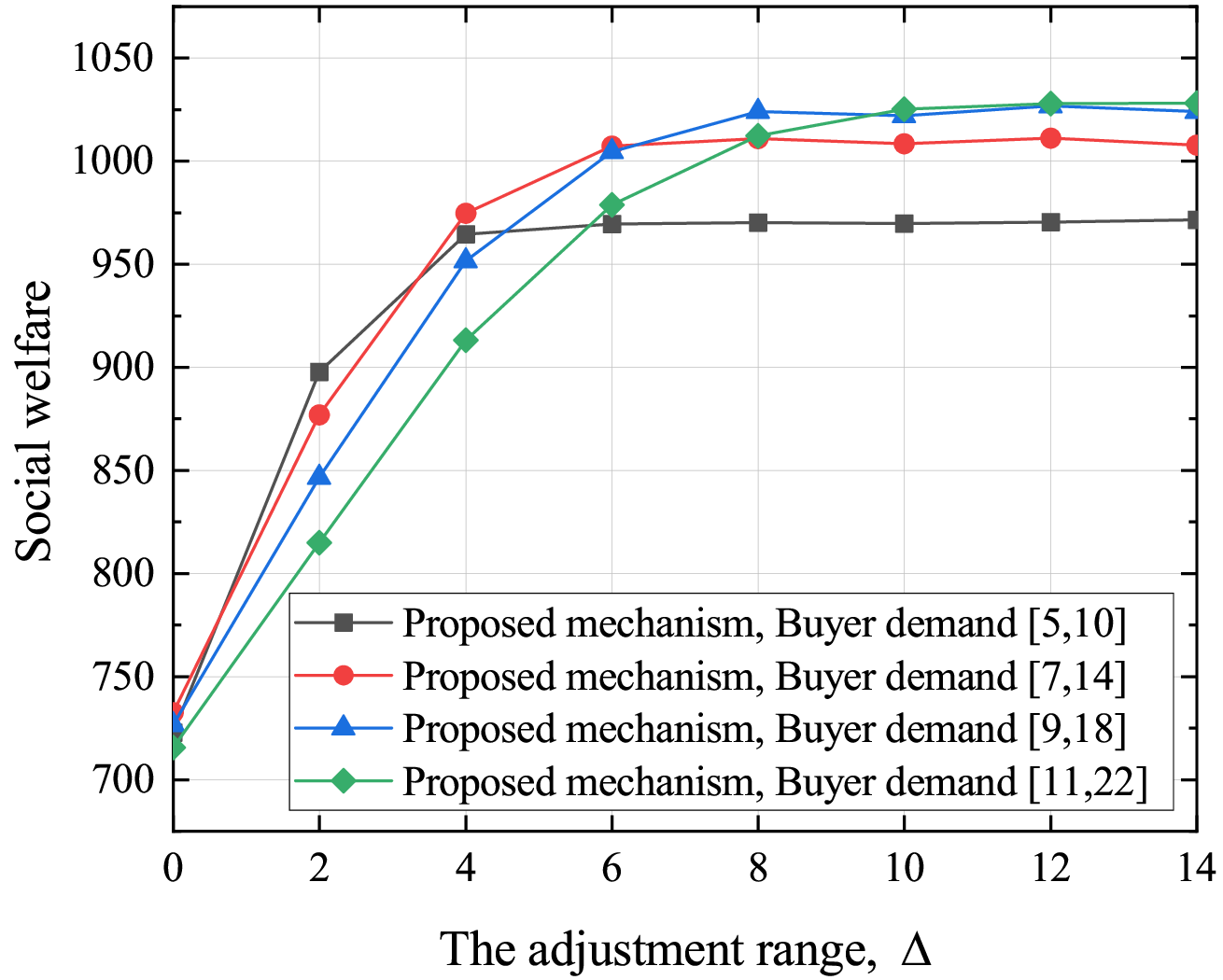}
	\caption{Social welfare comparison under different adjustment ranges and four spectrum demand distribution ranges, where  the number of buyer is 15.}
	\label{varying demand}
\end{figure}

Finally, we evaluate how varying adjustment ranges influence performance across diverse demand distributions.
Fig. \ref{varying demand} reveals two key insights: 1) For small values of $\Delta$, social welfare remains limited when the buyers' demand distributions become wider, due to insufficient flexibility; 2) However, as $\Delta$ approaches $10$, the mechanism reaches near-maximum flexibility, significantly improving social welfare, particularly for broader demand distributions. These results highlight the necessity of appropriately large adjustment ranges to effectively deal with diverse demand distributions, ensuring optimal social welfare outcomes.

\section{Conclusion}

To facilitate flexible and efficient spectrum sharing, this paper has proposed a novel approximately truthful combinatorial forward auction scheme with a flexible bidding mechanism. 
In the proposed scheme, buyers submit combinatorial bids that include both base spectrum demands and adjustable ranges, allowing for dynamic adaptation to market conditions.
We developed a GMWD mechanism, leveraging SEM coefficients to standardize valuations across heterogeneous frequency bands and optimize spectrum allocation.
Simulation results confirm that our approach consistently achieves higher social welfare compared to two benchmark methods, thereby validating the effectiveness and practicality of the proposed flexible bidding strategy.

\section*{Acknowledgement}
This work was supported in part by the National Natural Science Foundation of China under Grant 62371231, the Natural Science Foundation on Frontier Leading Technology Basic Research Project of Jiangsu under Grant BK20222001, and the Jiangsu Provincial Key Research and Development Program under Grants BE2023027.

%

\end{document}